\documentclass[aps,prl,twocolumn,superscriptaddress,amssymb]{revtex4}
\usepackage{epsfig}

\begin{document}

\title{
\vskip -50 pt
\begin{flushright}
\normalsize\rm NORDITA-2011-24
\end{flushright}
\vskip 20 pt
Dirac equation for membranes}

\author{M. Trzetrzelewski}
\email{maciej.trzetrzelewski@gmail.com}

\date{\today}

\setlength{\parindent}{2ex}
\setlength{\parskip}{1ex plus 0.5ex minus 0.2ex}

\begin{abstract}
 Dirac's idea of taking the square root of constraints  is
applied to the case of extended objects concentrating on membranes
in $D=4$ space-time dimensions. The resulting equation is Lorentz invariant and
predicts an infinite hierarchy of positive and negative masses
(tension). There are no tachyonic solutions.
\end{abstract}

\maketitle

\section{Motivation}

In the Dirac membrane model \cite{Dirac} the hamiltonian
corresponding to the spherically symmetric solution includes a
square root of the anharmonic oscillator. As noted by Dirac to
introduce the spin into the theory one would have to get rid of the
square root by bringing in the spin matrices. However, in the vast
literature on membranes (see e.g. \cite{Duff} and references
therein) a different way of introducing fermions for extended objects
is usually considered by writing the supersymmetric version of the
bosonic action \cite{supermembranes}. Besides aesthetic reasons this
seems quite not in the direction indicated by Dirac. It is perhaps
worth mentioning that for the case of point particles the
Klein-Gordon equation has its supersymmetric counterpart (i.e. the
massive Wess-Zumino model) which however does not describe the electron unlike the square root of the Klein-Gordon equation.

In this paper we choose a route indicated by Dirac i.e. introduce fermion
fields (and hence spin) for membranes by performing a certain square
root. We find it more appropriate to do so for the constraints of
the theory rather then for the hamiltonian since the former correspond to the mass shell constraint for point-like particles
and  are written in relativistically invariant way. 
However, taking the square root of the hamiltonian is also possible and was discussed in \cite{Stedile}.

The model presented here is manifestly Lorentz invariant hence there are
no anomalies originating form special relativity. There are also no
tachyonic solutions - we show that in the case of a spherical membrane the spectrum is real and discrete. The eigenvalues are identified with the membrane masses and can be positive as well as negative
corresponding to positive respectively negative membrane tension.
The field content (and their
equations) considered here is novel: the membrane is described by a fermionic field governed by the analog of the Dirac equation. It is an alternative way of describing extended objects in space-time.

Lastly we generalize the discussion by including a coupling to a
three form as well as suggest an alternative way of writing the Dirac
equation for membranes.

\section{A square root}
A Lorentz invariant action describing membranes is given by \cite{Dirac}
\begin{equation}  \label{mem}
S_{membrane}=-\Lambda\int \sqrt{G}d\sigma^3, \ \ \ G_{\alpha \beta}:=\partial_{\alpha} X_{\mu}\partial_{\beta} X^{\mu}
\end{equation}
where $\Lambda$ is the tension (here we assume $\Lambda>0$ and will comment on $\Lambda<0$ later),
$\sigma^{\alpha}=(\tau,\sigma^r)$ is the internal parametrization of
the membrane, $X^{\mu}$ are the embedding variables and $G$ is the
determinant of the induced metric $G_{\alpha\beta}$. The momenta $\mathcal{P}_{\mu}:=\partial
\mathcal{L}/\partial \dot{X}^{\mu}$ satisfy the constraints (see e.g. \cite{Barut})
\begin{equation} \label{cons2}
\mathcal{P}_{\mu}\mathcal{P}^{\mu}=\Lambda^2\det G_{rs}
\end{equation}
and follow from the diffeomorphism invariance of (\ref{mem}). In addition to (\ref{cons2}) we also have
$\mathcal{P}^{\mu}\partial_{r}X_{\mu} = 0$.

Equation (\ref{cons2}) is a counterpart of the mass shell constraint for a
point-like particle therefore in order to keep the analogy with the
way in which Dirac obtained equation for fermions it is desirable to
perform the square root of (\ref{cons2}).

\subsection{Spherically symmetric membrane}
Let us now consider a concrete example of a sphere at rest
\[
X^0=\tau, \ \  X^1=r(\tau)\sin\varphi \cos \theta, \ \ X^2=r(\tau)\sin\varphi \sin\theta,
\]
\[
X^3=r(\tau) \cos \varphi, \ \ \varphi \in [0,\pi], \ \ \ \ \theta \in [0,2\pi).
\]
The determinant in (\ref{cons2}) is $\det G_{rs}=r^4\sin^2\varphi$ hence
the corresponding square root is
\begin{equation} \label{diracmem0}
\gamma^{\mu}\mathcal{P}_{\mu}= -\Lambda A r^2\sin \varphi
\end{equation}
where $A$ is a matrix 
s.t. $A^2=\bold{1}$, $\gamma^{\mu}$ are gamma
matrices, $\{\gamma^{\mu},\gamma^{\nu}\}=2\eta^{\mu\nu}\bold{1}$, the minus sign is a convention. 

Equation (\ref{diracmem0}) is $\varphi$  dependent. A way to obtain the analog of the Dirac equation for the radial variable is to average
 (\ref{diracmem0}) over the membrane surface by
introducing the canonical momenta $\pi_{\mu}$ for the whole membrane via
\[
\pi_{\mu}:=\int_{\tau=const} \mathcal{P}_{\mu}d \varphi d \theta.
\]
This implies
\[
\gamma^{\mu}\pi_{\mu}= -M, \ \ \ M:=\Lambda r^2 \int A \sin\varphi d \varphi d \theta
\]
and suggests to consider the Dirac equation (substituting $\pi_{\mu}=-i \partial_{\mu}$) of the form
\begin{equation} \label{diracmemmass}
(-i\gamma^{\mu}\partial_{\mu} + M)\psi=0
\end{equation}
where $\psi=\psi(r)$. 

Because the spinor $\psi$ is independent of $\theta$ and $\varphi$ we may consider only the radial part of the differential operator in (\ref{diracmemmass}). The matrix $A$ should be chosen such that $M$ is proportional to the unit matrix. For $A=\bold{1}$ (other choices of $A$ are possible which after averaging lead to $M=0$), going to the radial equation (we use the conventions of \cite{Itzykson}) we obtain
\begin{equation} \label{free1}
\left(\partial_x + \frac{\kappa}{x}\right)G = ( \epsilon + x^2 )F, \ \ \ \left(-\partial_x + \frac{\kappa}{x}\right)F = ( \epsilon - x^2 )G
\end{equation}
where $x=r(4\pi\Lambda)^{1/3}$ and $\epsilon= E/(4\pi\Lambda)^{1/3}$ are dimensionless, $E$ is the energy, $\kappa=\pm 1$ and $F(x)$,
$G(x)$ are the radial parts of the spinor $\psi$  \cite{Itzykson}. Using matrix notation (\ref{free1}) can be written as
\[
H  \phi = \epsilon \phi, \ \ \  
H:= \left( \begin{array}{cc}
        -x^2 & \partial_x +\frac{\kappa}{x}     \\
        -\partial_x +\frac{\kappa}{x} &  x^2   \\
    \end{array} \right), \ \ \  
\phi := \left(\begin{array}{cc}
        F     \\
        G  \\
    \end{array} \right) . 
    \]
The spectrum of $H$ is independent of the choice of $\kappa$ (i.e. if $\phi^T=(F,G)$ solves $H\phi=\epsilon\phi$ for $\kappa=1$ then   $\phi^T=(G,F)$ solves $H\phi=\epsilon\phi$ for $\kappa=-1$) hence we choose $\kappa=1$ from now on.   
Let us note that if $\phi$ is the eigenvector of $H$ then (due to the $1/x$ term) $F(0)=G(0)=0$. Therefore we may consider the Hilbert space of square-integrable vectors $\phi$ on $[0,\infty)$ satisfying $\phi(0)=0$. In such Hilbert space the operator $\partial_x$ is antihermitian hence $H$ is hermitian which proves that the spectrum of $H$ is real. 

Let us also observe that $H^2$ can be written as
\[
H^2=Q^2+ 
\left( \begin{array}{cc}
        0 & 0     \\
        0  &  2/x^2   \\
    \end{array} \right), \ \ \ 
    Q:= \left( \begin{array}{cc}
        -x^2 & \partial_x     \\
        -\partial_x  &  x^2   \\
    \end{array} \right)
\]
hence the spectrum of $H^2$ is discrete, due to the inequality $H^2 \ge Q^2$ and the fact that $Q^2$ is discrete (to see explicitly that $Q^2$ is discrete it is useful to introduce  $a_{\pm}:=F\pm G$ for which the eigen-equation $Q^2\phi=\eta^2 \phi$ gives
$
h_{\pm}a_{\pm}= \eta^2 a_{\pm}, \ \ \  h_{\pm}:=-\partial_x^2 + x^4 \pm 2x,
$
i.e. $a_{+}$ and $a_{-}$ decouple and since $h_{+}$and $h_{-}$ are discrete, $Q^2$ must also be discrete). This proves that the spectrum of $H^2$ and hence $H$ is discrete. 

To find the exact spectrum of $H$ we use numerical methods (see the Appendix for details). The first positive  energy levels are
\[
\epsilon \approx   2.7, \ 4.0, \ 5.1, \  6.1, 7.0, \ 7.9, \ 8.7, \  9.5, \ \ldots 
\]
while the negative ones are
\[
-\epsilon \approx   1.7, \ 3.3, \  4.5, \  5.6, \ 6.6, \ 7.5, \ 8.3, \ 9.1, \   \ldots \ .
\]

 The four degrees of
freedom of the spinor $\psi$ should  be interpreted just like in the
Dirac equation i.e. particles and antiparticles with spin (the notion of spin is not related here to any geometrical configuration such as e.g. the spinning membrane).  The
antiparticle corresponds to the negative energy $-E$ however the
energy in this membrane model is a mass. Therefore one concludes
that the model predicts the existence of particles with negative
masses. Because the energy is measured in units of $\Lambda^{1/3}$
and since  $(-\Lambda)^{1/3}=-\Lambda^{1/3}$ we can identify the
negative masses with the negative tension - a possibility already
considered by Dirac (see the second reference in \cite{Dirac}). Such
objects are unstable (unless one considers gravitational effects
with extra volume term in the action \cite{Dirac}). The inclusion of
the electromagnetic field would only worsen the stability of
particles with negative tension.

The electromagnetic field can be added using minimal coupling prescription $\mathcal{P}_{\mu} \to  \mathcal{P}_{\mu}  + e A_{\mu}$ with $A_{0}=-e/2r$ (the electrostatic energy of the Coulomb field of a charged sphere) and $A_i=0$. Equations (\ref{free1}) stay the same with shifted energy $\epsilon \to \epsilon +\frac{\alpha}{2x}$ where $\alpha$ is the fine structure constant. The corresponding hamiltonian is
\[
H^{EM}=H+\frac{\alpha}{2x}\bold{1}.
\]
Due to the small value of $\alpha$ the spectrum of $H^{EM}$ is very close to the spectrum of $H$ (differences are of order $10^{-3}$, the lowest positive energy is now 2.7570 compared to 2.7525 obtained previously). 

Let us compare the spectrum of $H^{EM}$ with the spectrum of the hamiltonian for the bosonic membrane obtained by Dirac \cite{Dirac}
\begin{equation} \label{1}
H_{Dirac}=\sqrt{-\partial_r^2+\omega^2 r^4 } +\frac{e^2}{2r},
\end{equation}
where $\omega=4\pi \Lambda$. Using the Bohr-Sommerfeld quantization method, Dirac found
that the first excitation of the membrane corresponds to the energy $\approx 53m_e$, where $m_e$ is the mass of the electron. The exact value turns out to be about $ 43m_e$ as observed in \cite{gnadig}. Using the relation $\omega=16 m_e^3/27 \alpha^2$ \cite{Dirac} we find that our model gives
\[
E= (4\pi \Lambda)^{1/3} \epsilon_0 \approx  \frac{1}{3} \left(\frac{4}{\alpha}\right)^{2/3}m_e \cdot 2.757 = 61.5 m_e.
\]
Therefore we obtained a larger value - in accordance with Dirac's expectation that bringing spin into the theory might lift the eigenvalue. However the result is still not of order of the muon mass.

\subsection{General case}

Taking the square root of (\ref{cons2}) is more difficult.
While the l.h.s.  would simply give
$\gamma^{\mu}\mathcal{P}_{\mu}$  the r.h.s. involves a square root of the determinant $\det
G_{rs}$ which is 4th order in derivatives. It turns out that the
answer is simple in terms of a hermitian mass matrix density
\begin{equation} \label{mass}
\mathcal{M}:=\frac{1}{2}i\Lambda \epsilon^{rs}\gamma_{\mu\nu}\partial_r X^{\mu}\partial_s X^{\nu}
\end{equation}
(where
$\gamma_{\mu\nu}=\gamma^{\rho\sigma}\eta_{\rho\mu}\eta_{\sigma\nu}$,
$\gamma^{\mu\nu}:=\frac{1}{2}[\gamma^{\mu},\gamma^{\nu}]$,
$\epsilon^{rs}$  is completely antisymmetric with $\epsilon^{12}=1$)
due to the identity

\begin{equation} \label{identity}
\mathcal{M}^2=\Lambda^2\det G_{rs}\bold{1}
\end{equation}
which allows to write the square root of the constraint as
\begin{equation} \label{diracmem}
\gamma^{\mu}\mathcal{P}_{\mu}=-\mathcal{M}.
\end{equation}
It is possible to introduce a matrix $\mathcal{A}$ such that the rescaling $\mathcal{M_A} := \mathcal{A}  \cdot \mathcal{M}$ is consistent with the constraint (\ref{cons2}). One simple example is $\mathcal{A}=A\mathcal{M}/\Lambda\sqrt{\det G_{rs}}$
where matrix $A$ as s.t. $A^2=\bold{1}$, for which $\mathcal{M}_\mathcal{A}=\Lambda A \sqrt{\det G_{rs}} \bold{1}$. Therefore this particular choice of $\mathcal{A}$ will reproduce the results of previous section.

The choice  $\mathcal{A}=\bold{1}$ results in  the averaged mass matrix $M$ equal $0$ for spherical membrane, hence the spectrum of the corresponding Dirac operator would be continuous. We now elaborate on this case in more details since it is where quantization procedure is possible in terms of matrix regularization \cite{hoppephd}. 

If the parameters $\sigma^r$ are not
integrated out the
coordinates $X^{\mu}$ are fields and the differential equation of
the quantum counterpart of (\ref{diracmem}) is a functional one.
Because $\mathcal{P}_{\mu}$ and $X^{\mu}$ are conjugate variables it is
natural to substitute $\mathcal{P}_{\mu}$ with a functional derivative
$-i\frac{\delta}{\delta X^{\mu}}$ in (\ref{diracmem}). A more
problematic term is the mass matrix density $\mathcal{M}$. Observing
the appearance of the Poisson bracket
$\{X^{\mu},X^{\nu}\}=\epsilon^{rs}\partial_r X^{\mu} \partial_s
X^{\nu} $ in (\ref{mass}) we write
\begin{equation} \label{diracmem3}
\left(-i\gamma^{\mu}\frac{\delta}{\delta X^{\mu}} + \frac{1}{2}i\Lambda \gamma_{\mu\nu}\{X^{\mu},X^{\nu}\}  \right) \Psi=0
\end{equation}
which will be useful when using the matrix  regularization. The above equation is similar to Eq. (47) of Ref. \cite{Smolin} where an attempt to quantize the bosonic membrane in a covariant fashion was undertaken.

Equation (\ref{diracmem3}) at least formally is solved by
\begin{equation} \label{solution}
\Psi = e^{S}\Psi_0, \ \ \ S=\frac{\Lambda}{12}\gamma_{\mu\nu\rho}\int X^{\mu}\{X^{\nu},X^{\rho}\} d^3\sigma
\end{equation}
where $\Psi_0$ is a constant spinor,
$\gamma_{\mu\nu\rho}=\frac{1}{3}(\gamma_{\mu}\gamma_{\nu\rho}+cycl.)$
and where we used
$\gamma^{\mu}\gamma_{\mu\nu\rho}=2\gamma_{\nu\rho}$. Similar
functional expressions were found for the bosonic membrane by Smolin
\cite{Smolin}, Moncrief \cite{Moncrief} and Hoppe \cite{Hoppe}.  Using the identities
$\gamma_{\mu\nu\rho}^{\dagger}=-\gamma_0 \gamma_{\mu\nu\rho}\gamma_0$ and  $e^{S^{\dagger}}=
\gamma_0  e^{-S} \gamma_0$ it follows that $\bar{\Psi}\Psi = \bar{\Psi}_0\Psi_0 $ therefore $\Psi$ suffers from being non-normalizable (i.e. the functional integral $\int [dX] \bar{\Psi}\Psi $ is infinite).

The hamiltonian formulation can be obtained by singling out the
hamiltonian operator $\mathcal{H}:=\mathcal{P}_0$ and noting that the gauge choice
$X^0=\sigma^0$ for the coordinates implies that $\{X_{0},X_{\nu}\}=0$ hence
\begin{equation} \label{hamgen}
\mathcal{H}\Psi = E\Psi, \ \ \ \ \mathcal{H}=\gamma^0\left(\gamma^{i}\mathcal{P}_{i} + \frac{1}{2}\Lambda \gamma_{ij}\{X^i,X^j\}  \right).
\end{equation}
In addition to (\ref{diracmem3}) we also have the constraint $\mathcal{P}_{\mu}\partial_r X^{\mu}=0$ which after setting the gauge $X^0=\sigma^0$ becomes $\mathcal{P}_i\partial_r X^i=0$. Locally (on the membrane surface) this constraint is equivalent to $\{\mathcal{P}_i,X^i\}=0$ which will be further used.
One can write a solution analogous to (\ref{solution}) corresponding to $E=0$
\begin{equation} \label{solutionh}
\Psi = e^{S}\Psi_0, \ \ \ S=\frac{\Lambda}{6}\gamma_{ijk}\int X^{i}\{X^{j},X^{k}\}d^2\sigma
\end{equation}
(where we used
$\gamma^{i}\gamma_{ijk}=\gamma_{jk}$) which however is not normalizable by similar argument as before.

Using matrix regularization \cite{hoppephd} we can view (\ref{hamgen}) as the $N \to \infty$ limit of
\[
H\Psi = E\Psi, \ \ \ \ H=\gamma^0\gamma^{i}p_{i} + \frac{1}{2}\Lambda \gamma^0\gamma_{ij}[x^i,x^j] 
\]
where $p_i$ and $x^i$ are $su(N)$ matrices, subject to the constraint $[p_i,x^i]\Psi=0$. 
The solution corresponding to $E=0$ is
\begin{equation} \label{solutionhreg}
\Psi = e^{S}\Psi_0, \ \ \ S=\frac{\Lambda}{6\sqrt{N^2-1}}\gamma_{ijk}Tr(x^i[x^j,x^k])
\end{equation}
(using the conventions of \cite{HoppeMT}). The results of the previous section suggest that the spectrum of membrane excitations does not contain the state with $E=0$ therefore one should not worry that the wavefunctions (\ref{solutionh}) and (\ref{solutionhreg}) are not normalizable. However the exponential factor $e^S$ will probably play an important role in search for normalizable excitations. 
In the matrix-regularized case this would imply that the tension has to be renormalized as $\Lambda \sim N$.

The electromagnetic field  $A_{\mu}$ can be introduced using the
minimal coupling prescription $-i\frac{\delta}{\delta X^{\mu}} \to
-i\frac{\delta}{\delta X^{\mu}} +e A_{\mu}$. How one can write the
matrix regularized case is less obvious since $A_{\mu}$ are in
general non polynomial.

\section{3-form coupling}
 Let us consider a different approach based on $p$-form electrodynamics \cite{Henneaux} which is
more natural for extended objects with $p-1$ spatial directions. We
therefore assume the existence of a three form
$\mathcal{A}$ in space-time, with components given by a completely antisymmetric
tensor $\mathcal{A}_{\mu\nu\rho}$ and the corresponding field
strength $\mathcal{F}_{\mu\nu\rho\sigma}=\partial_{[\mu}\mathcal{A}_{\nu\rho\sigma]}$

For the action of the membrane coupled with $\mathcal{A}$ we take
\begin{equation} \label{EMaction}
S=-\Lambda\int \sqrt{G}d^3\sigma -\frac{e}{3!}\int \epsilon^{\alpha\beta\gamma}\mathcal{A}_{\mu\nu\rho}\partial_{\alpha} X^{\mu}\partial_{\beta} X^{\nu}\partial_{\gamma} X^{\rho} d^3\sigma
\end{equation}
so that the momenta and the equations of motion are
\[
\mathcal{P}_{\mu}=-\Lambda\sqrt{|G|}G^{\tau \beta}\partial_{\beta} X_{\mu}-\frac{1}{2}e \mathcal{A}_{\mu\nu\rho}\{X^{\nu},X^{\rho}\},
\]
\[
\Lambda\partial_\alpha \sqrt{G}G^{\alpha\beta}\partial_{\beta}X^{\mu}+e{\mathcal{F}^{\mu}}_{\nu\rho\sigma}\{X^{\nu},X^{\rho},X^{\sigma}\}=0
\]
where
$\{X^{\mu},X^{\nu},X^{\rho}\}:=\epsilon^{\alpha\beta\gamma}\partial_{\alpha}X^{\mu}\partial_{\beta}X^{\nu}\partial_{\gamma}X^{\rho}$
is the Nambu bracket \cite{Nambubra}. The corresponding square root
of constraints will result in (\ref{diracmem3}) with the functional
derivative $-i\frac{\delta}{\delta X^{\mu}} \to
-i\frac{\delta}{\delta X^{\mu}} +\frac{1}{2}e
A_{\mu\nu\rho}\{X^{\nu},X^{\rho}\}$. One can also write the matrix
regularization.

Let us note that the
equations of motion for $\mathcal{A}_{\mu\nu\rho}$ do not contradict
sourceless Maxwell equations in the following sense: using
the identity
$\mathcal{F}_{\mu\nu\rho\sigma}=-6\epsilon_{\mu\nu\rho\sigma}\partial_{\pi}a^{\pi}$
where $a^{\pi}:=\frac{1}{6}\epsilon^{\pi\mu\nu\rho}\mathcal{A}_{\mu\nu\rho}$
($\epsilon_{\mu\nu\rho\sigma}$ is completely antisymmetric,
$\epsilon_{1234}=1$), the action for $\mathcal{F}$ becomes
\[
S_{\mathcal{F}}= -\frac{1}{2\cdot4!}\int\mathcal{F}_{\mu\nu\rho\sigma}\mathcal{F}^{\mu\nu\rho\sigma}d^4x = 18\int(\partial_{\mu}a^{\mu})^2d^4x
\]
i.e. a square of the Lorenz gauge condition. The
equations of motion $\partial_{\nu}(\partial_{\mu}a^{\mu})=0$ imply
$\partial_{\nu}a^{\nu} = const.$ hence the corresponding space of
solutions is larger then that of the electromagnetic potential
$A_{\mu}$ in the Lorenz gauge. Moreover the residual gauge
invariance of $\mathcal{F}^2$ is present with  $a_{\mu}\to
a_{\mu}+\partial_{\mu}\chi$, $\Box\chi=0$ just like in the
formulation of electrodynamics in the Lorenz gauge.

\section{Alternative square root}
An identity similar to (\ref{identity}) holds also for the whole world-volume
metric i.e.
\[
\mathcal{L}^2=\det G_{\alpha\beta}\bold{1}, \ \ \ \ \mathcal{L}:=\frac{i}{3!} \gamma_{\mu\nu\rho}\{ X^{\mu}, X^{\nu}, X^{\rho}\}.
\]
Both identities have also a heuristic origin when introducing a
matrix line element $dX:=\gamma_{\mu}dx^{\mu}$ and noting that
components of the matrix surface and volume elements are
\[
dX \wedge dX = \frac{\gamma_{\mu\nu}}{2!}\{ X^{\mu},X^{\nu}\} d\sigma^1 \wedge d\sigma^2
\]
for one forms $dx^{\mu}=\partial_r X^{\mu} d\sigma^r$ on the surface and
\[
dX \wedge dX \wedge dX=\frac{\gamma_{\mu\nu\rho}}{3!}\{ X^{\mu}, X^{\nu}, X^{\rho}\}d\sigma^0 \wedge d\sigma^1 \wedge d\sigma^2
\]
for one forms $dx^{\mu}=\partial_{\alpha} X^{\mu} d\sigma^{\alpha}$ on the
world-volume. Therefore matrices $\mathcal{M}$ and $\mathcal{L}$
appear quite naturally when working with matrix line element $dX$.

Having in mind the above remarks let us consider the following analogy. In the case of point particles the Lagrangian formally can be written as a matrix
\[
\sqrt{\dot{x}_{\mu}\dot{x}^{\mu}}d\tau = \gamma_{\mu}\dot{x}^{\mu}d\tau=\gamma_{\mu}dx^{\mu}
\]
which has a counterpart in $-i\gamma^{\mu}\partial_{\mu}$ - the
differential part of the Dirac operator. Likewise, for membranes we have
\[
\sqrt{G}d^3\sigma=\frac{i}{6} \gamma_{\mu\nu\rho}\{ X^{\mu}, X^{\nu}, X^{\rho}\}d^3\sigma=i\gamma_{\mu\nu\rho}d\Omega^{\mu\nu\rho}
\]
where $d\Omega^{\mu\nu\rho}:=\frac{1}{6} \{ X^{\mu}, X^{\nu}, X^{\rho}\}d^3\sigma$ is the element of the world-volume, hence the corresponding Dirac operator would be $\gamma^{\mu\nu\rho}\frac{\delta}{\delta\Omega^{\mu\nu\rho}},$
with the functional derivative (already considered by Nambu
\cite{Henneaux}). Therefore in the presence of the field
$\mathcal{A}_{\mu\nu\rho}$ we would obtain
\[
\gamma^{\mu\nu\rho}\left(-i\frac{\delta}{\delta\Omega^{\mu\nu\rho}}+e\mathcal{A}_{\mu\nu\rho}\right)\Psi=0
\]
which is an alternative way to write the Dirac equation with the three-form.

\section{Conclusions}

A theory of extended objects aspiring to
describe particles should include fermionic fields. How to write
down such theory is non-trivial considering the fact that the
starting point is a classical object described by bosonic fields
(coordinates). A problem of this kind but for point-like particles
was solved by Dirac by taking a square root of the constraints and since
it is the correct route we take it as a guiding principle in the
attempt to include fermions into the theory of extended objects.

Because the constraints are local equations depending on the
geometry of the membrane, the corresponding Dirac equation will be a
functional one. Even for the simplest choice of the matrix $\mathcal{A}$, this results in a highly
complicated system (\ref{diracmem3}) which fortunately can be
understood using the matrix regularization so that the masses of the
membrane are hidden in the spectral properties of such  matrix-Dirac
operator. Still to find the eigenvalues of this equation is probably
very difficult which is why we considered a semiclassical approach
by concentrating on spherically symmetric membranes. The resulting
system is a Dirac equation for the radius of the membrane. Although
this approach is only approximate we believe the we have identified
the crucial properties of the theory i.e. that the spectrum is
discrete and consists of positive as well as negative eigenvalues - in particular there is no 0 eigenvalue and  there are no tachyons.
The negative masses are due to the negative tension and therefore
such particles would be unstable.

\vspace{6pt} \noindent{\bf Acknowledgments} Discussions and the
correspondence with J. Hoppe, M. Ku\'zniak and P. O. Mazur as well
as the support from the Swedish Research Council, KTH and NORDITA are
gratefully acknowledged.

\end{document}